\documentclass[preprint,12pt]{elsarticle}
\usepackage{dcolumn}%
\usepackage{bm}%
\usepackage{graphicx}
\usepackage{amsfonts}
\usepackage{amsmath}
\usepackage{amssymb}
\usepackage{hyperref}
\usepackage{indentfirst}
\usepackage[nodots]{numcompress}
\usepackage[usenames,dvipsnames]{pstricks}
\usepackage{epsfig}
\usepackage{pst-grad} % For gradients
\usepackage{pst-plot} % For axes
\journal{Physics Letters A}

\begin{document}
\begin{frontmatter}

\title{On geometry-dependent vortex stability and topological spin excitations on curved surfaces with cylindrical symmetry}
\author[1]{V. L. Carvalho-Santos}
\ead[1]{vagson.santos@bonfim.ifbaiano.edu.br}
\address[1]{Instituto Federal de Educa\c c\~ao, Ci\^encia e Tecnologia Baiano - Senhor do Bonfim\\48970-000 Senhor do Bonfim, Bahia, Brazil}
\author[2]{F. A. Apol\^onio}
\address[2]{Departamento de F\'isica, Universidade Federal de Vi\c cosa\\36570-000 Vi\c cosa, Minas Gerais, Brazil}
\author[3]{N. M. Oliveira-Neto}
\address[3]{Departamento de Qu\'imica e Exatas, Universidade Estadual do Sudoeste da Bahia\\ 45206-190, Jequi\'e, Bahia, Brazil.}

\begin{abstract}We study the Heisenberg Model on cylindrically symmetric curved surfaces. Two kinds of excitations are considered. The first is given by the isotropic regime, yielding the sine-Gordon equation and $\pi$-solitons are predicted. The second one is given by the XY model, leading to a vortex turning around the surface. Helical states are also considered, however, topological arguments can not be used to ensure its stability. The energy and the anisotropy parameter which stabilizes the vortex state are explicitly calculated for two surfaces: catenoid and hyperboloid. The results show that the anisotropy and the vortex energy depends on the underlying geometry.
\end{abstract}
\end{frontmatter}

% \linenumbers

\section{Introduction}
Geometrical and topological concepts and tools are important in many branches of natural sciences, particularly, in Physics. For instance, the idea of symmetry, which is intimately associated with geometry, is a keystone for studying a number of fundamental properties of several physical systems, e.g., the Noether theorem asserts that there is a conserved quantity to each continuous symmetry of the associated action. Topology, in turn, is crucial for classifying and  for giving stability to certain excitations, such as solitons, extending objects having finite energy, and vortices, presenting a non-vanishing vorticity around a given singular point or a topological obstruction. In addition, the observed vortex-pair dissociation is the mechanism behind the topological phase transition \cite{kosterlitz-jphysC}. Vortices and solitons have been observed in a number of systems, such as superconductors, superfluids, and magnetic materials \cite{vortexobservation,solitonobservation}.

Curvature effects play an important role in the characteristics of these topological structures. For instance, Vitelli \textit{et al} have shown that in-plane vortices interact not only with each other, but also with the curvature of the substrate \cite{Vitelli-PRL93}. Curvature is also an important factor in the magnetic systems behaviour, in which the interaction of the out-of-plane component of magnetic vortices with curved defects must cause a chiral symmetry breaking in its gyrotropic motion due the thin-film roughness \cite{Vansteenkiste,apolonio}. Furthermore, the easy-surface Heisenberg model in magnetic spherical shells predicts a coupling between the localized out-of-surface component of the vortex with its non-localized in-surface structure, associated with the curvature of the underlying geometry \cite{Kravchuk-Arxiv} and still, the smooth and variable curvature of ferromagnetic nanotorus ensures the stability of the vortex for smaller radius than their nanoring counterparts \cite{Vagson-JAP}.

In the case of two-dimensional systems, vortices and solitons can appear like solutions of the continuous Heisenberg Model, which has been used to analyse the dynamic and static properties of vortices, showing that the energy of these excitations is closely linked to the geometrical properties of the surface \cite{torusmeu}$-$\cite{cone}. It has also been shown that, for simply-connected surfaces, the vortex energy presents a divergence, which can be controlled by the insertion of a cutoff in the region where the continuous limit of Heisenberg Hamiltonian has not validity. In the case of magnetic systems, this divergence must be controlled by the development of an out-of-plane component in the vortex core region. Soliton-like solution has also been considered in the above cited works and it has been shown that its characteristic length depends on the length scale of the surface. For finite surfaces, fractional/half-soliton solutions have been found \cite{pseudosphere,Saxena-PRB66}. Furthermore, the interaction of an external magnetic field with Heisenberg spins on a cylindrical surface yields a 2$\pi$ soliton-like solution, inducing a deformation at the sector where the spins are pointing in the opposite direction to the magnetic field \cite{Saxena-PRB58}. A 2$\pi$ soliton has also been predicted to appear on curved surfaces with cylindrical symmetry, provided the magnetic field is coupled with the curvature of the substrate \cite{Vagson-PLA,Vagson-BJP}.

In this paper, we study the anisotropic Heisenberg Model on curved surfaces with cylindrical symmetry. We are interested in studying a class of topological spin textures on these manifolds in such way that both, soliton and vortex-like solutions are considered. Solitons are predicted to appear on an infinite cylindrically symmetric surface, if the isotropic case is taken into account. In our assumptions, the soliton characteristic length is rescheduled to one and does not depend on the characteristic length scale of the surface. For finite surfaces, fractional solitons, which have not topological stability, are found. 

Our analysis includes the study of the XY model and vortex-like solutions are considered. It is shown that, for non-simply connected manifolds, the obstruction of the surface ensures the vortex topological stability due the removal of spurious divergences appearing in the core region. The XY model can also yield a helical-like state, however topological arguments must not be used to ensure the stability of this spin configuration. The energies of these spin textures are calculated and we we get that it depends on the surface curvature. Furthermore, we calculate the critical anisotropy parameter for which the vortex appears as the ground state and we show that it is also associated to the geometrical properties of the surface. We calculate explicitly the vortex energy and the critical anisotropy parameter for two different geometries: the catenoid and hyperboloid, which are negatively curved and non-simply connected manifolds. The choice for studying these surfaces is associated with the fact that both can be realized in fluid interfaces provided with an orientational ordered phase as a consequence of the interplay between surface tension and orientational elasticity \cite{Giomi-PRL-109}. Furthermore, catenoid is the shape minimizing the curvature elastic energy, appearing in phospholipid vesicles of high topology \cite{michalet-PRL-72} and the symmetry of these geometries allows us to compare our results to these found for the cylindrical case, largely explored in nanomagnetism researches.

To proceed with our analysis, this work is organized as follows: in Section \ref{HeisModel} we present the continuous anisotropic Heisenberg Model on rotationally symmetric surfaces. The results and discussions for the isotropic Heisenberg Hamiltonian and the XY model are also considered in this section. Section \ref{partcases} brings the discussions about the model on the catenoid and hyperboloid surfaces and compare our results with that obtained for the surface of a cylinder. Finally, in the Section \ref{conclusion}, we present our conclusions and prospects for future works.

%===============================================================================================
%===============================================================================================
%===============================================================================================

\section{Continuum Heisenberg Model on Curved Surfaces}\label{HeisModel}

The anisotropic exchange Heisenberg model, for nearest neighbour interacting spins on a two-dimensional lattice, is given by the Hamiltonian below:
\begin{equation}
\label{heisdisc} {H}_{\text{latt}}=-J'\sum_{\langle i,j\rangle}
[m_{i}^{x}m_{j}^{x}+m_{i}^{y}m_{j}^{y}+(1+\lambda)m_{i}^{z}m_{j}^{z}], 
\end{equation}
where $J'$ denotes the coupling between neighbouring spins, and according to $J'<0$ or $J'>0$, the Hamiltonian describes a ferro or antiferromagnetic system, respectively. $\vec{m_{i}}=(m_{i}^{x},m_{i}^{y},m_{i}^{z})$ is the spin operator at site $i$ and the parameter $\lambda$ accounts for the anisotropy interaction amongst spins: for $\lambda>0$, spins tend to align along the internal $Z$ axis (easy-axis regime); for $\lambda=0$, one gets the isotropic case; for $-1<\lambda<0$, we have the easy-plane regime, while the $\lambda=-1$ case yields to the so-called XY model, which has been considered on curved surfaces \cite{XY-model}. If we focus on a two-component spin, imposing $m_{z}\equiv0$, so that $\vec{m}_{PRM}=(m_{x},m_{y})$, we get the planar rotator model (PRM).

In the continuum approach of spatial and spin variables, valid at sufficiently large wavelength and low temperature, the model given by Eq. (\ref{heisdisc}) may be written as follows $(J\equiv J'/2)$:
$$H=-\frac{4J}{a^2}\iint \frac{1}{\sqrt{|g|}}(1+\lambda m_z^2)d\eta_1d\eta_2$$
\begin{equation}
\label{heiscont} 
+J\iint\sum_{i,j=1}^{2}\sum_{a,b=1}^{3}
g^{ij}h_{ab}(1+\delta_{a3}\lambda)
\left(\frac{\partial
m^{a}}{\partial\eta_{i}}\right)\left(\frac{\partial
m^{b}}{\partial\eta_{j}}\right)\sqrt{|g|}d\eta_{1}d\eta_{2},
\end{equation} 
where $a$ is the network spacing, the surface has curvilinear coordinates $\eta_{1}$ and $\eta_{2}$, $\sqrt{|g|}=\sqrt{|det[g_{ij}]|}$, $g^{ij}$ and $h_{ab}$ are the surface and spin space metrics, respectively (as usual, $g^{ij}g_{jk}=\delta^{i}_{k}$). Now, $\vec{m}=(m_{x},m_{y},m_{z})\equiv(\sin\Theta\cos\Phi,\sin\Theta\sin\Phi,\cos\Theta)$
is the classical spin vector field valued on an unity sphere (internal space), so that $\Theta=\Theta(\eta_{1},\eta_{2})$ and $\Phi=\Phi(\eta_{1},\eta_{2})$. With this, the Cartesian parametrization for $\vec{m}$ yields to $h_{ab}=\delta_{ab}$. Note that the first term in the first integral in the Hamiltonian (\ref{heiscont}) is the ground state energy and we will renormalize it to be zero. It can also be noted that, if $\lambda$ decreases from 0 to $-1$, the term depending on $m_z^2$ increases the energy if $m_z\neq0$, thus, the smallest energy associated to the anisotropic Heisenberg Hamiltonian will occur for $m_z=0$. The Hamiltonian (\ref{heiscont}) may be also viewed as the anisotropic non-linear $\sigma$ model (NL$\sigma$M), which lies on an arbitrary two-dimensional geometry. Thus, besides ordinary spins, the above model can be used to describe another condensed matter systems, e.g., a superfluid helium film, thin superconducting films \cite{Vitelli-PRL93,Mermin-Review}, a nematic liquid crystal confined on curved surfaces \cite{Napoli-PRL-108} or a spin lader, which consists in two or more coupled spin chains \cite{Pereira-SSC}. 

Our interest is to study the above model on curved surfaces with cylindrical symmetry, which, in cylindrical coordinate system, can be parametrized by $\mathbf{r}=(\rho(z),\phi,z)$, where $\rho(z)\equiv\rho$ is the radius of the surface at height $z$, and $\phi$ accounts for the azimuthal angle. In this case, we have that the covariant metric elements are given by:
\begin{equation}
g_{\phi\phi}=\frac{1}{g^{\phi\phi}}=\rho^2\hspace{1cm}\text{and}\hspace{1cm} g_{zz}=\frac{1}{g^{zz}}=1+\rho\,'\,^2,
\end{equation}
where $\rho\,'=d\rho/dz$. In this case, the Hamiltonian (\ref{heiscont}) can be rewritten as:
$$H=J\iint\Big\{\sqrt{\frac{g_{\phi\phi}}{g_{zz}}}\left[f(\Theta)(\partial_{z}
\Theta)^2+\sin^2\Theta(\partial_{z}\Phi)^2\right]$$
\begin{equation}\label{HamGen}
+\sqrt{\frac{g_{zz}}{g_{\phi\phi}}}\left[f(\Theta)(\partial_{\phi}
\Theta)^2+\sin^{2}\Theta(\partial_{\phi}\Phi)^2\right]-\sqrt{\frac{g^{zz}}{g_{\phi\phi}}}\frac{4\lambda\cos^2\Theta}{a^2}\Big\}dz d\phi,
\end{equation}
where $f(\Theta)=1+\lambda\sin^2\Theta$.

The Euler-Lagrange equations derived from the Hamiltonian (\ref{HamGen}) are evaluated to give:
$$2f(\Theta)(\partial^2_{\zeta}\Theta+\partial^2_{\phi}\Theta)=\sin2\Theta\Big\{\left(\partial_{\zeta}
\Phi\right)^2+\left(\partial_{\phi}\Phi\right)^2$$
\begin{equation}\label{GenForm}
-\lambda\left[\left(\partial_{\zeta}
\Theta\right)^2+\left(\partial_{\phi}\Theta\right)^2\right]+\frac{4\lambda}{a^2}\sqrt{\frac{g^{zz}}{g_{\phi\phi}}}\Big\}
\end{equation} 
and
\begin{equation}\label{phieq}
\sin^2\Theta\left(\partial^{2}_{\zeta}\Phi+\partial^{2}_{\phi}\Phi\right)+\sin2\Theta\left(
\partial_{\zeta}\Theta\partial_\zeta\Phi+\partial_{\phi}\Theta\partial_\phi\Phi\right)=0,
\end{equation}
where $d\zeta=\sqrt{{g_{zz}}/{g_{\phi\phi}}}dz$.

At this point, one can note that the above equations resemble, in form, their counterparts for the planar, cylindrical \cite{cylinder}, spherical \cite{sphere}, pseudospherical \cite{pseudosphere} and toroidal \cite{torusmeu} surfaces. Indeed, whenever $\zeta$ is identified with 
$$\zeta_{\text{cyl}}=\int\frac{1}{\rho}d\rho,\hspace{1cm}
\zeta_{\text{tor}}=\int\frac{s}{\mathcal{R}+s\sin\theta}d\theta,$$ 
 \begin{equation}\label{ZetaPar}
\zeta_{\text{sph}}=\int\frac{1}{\mathcal{S}\sin\theta} d\theta,\hspace{1cm}\zeta_{\text{psph}}=\int\frac{1}{\sinh\tau}d\tau,
\end{equation} 
the Eqs. (\ref{GenForm}) and (\ref{phieq}) recover their cylindrical, toroidal, spherical or pseudospherical analogs, respectively. Here, $\theta$ is the polar angle in spherical coordinates system. In the toroidal surface, it plays a similar role, however $\theta$ accounts for the polar angle between the spin vector and a line parallel to the $z$-axis, crossing the center of the torus arm. $\mathcal{R}$ and $s$ are the rotating and axial radius of the torus, $\mathcal{S}$ is the sphere radius and $\sinh\tau$ accounts for the distance measured along pseudospherical geodesic, say, a hyperbole. 

As expected, the anisotropic Heisenberg model is described by nonlinear differential
equations and suitable nontrivial solutions can be obtained provided some conditions are imposed. Thus, special solutions for the most general Eq. (\ref{GenForm}) will be explicitly obtained by solving it for particular values of $\lambda$. Initially, we will consider the isotropic case, which is given by $\lambda=0$. After, we will take $\lambda=-1$ to study the XY model.

\subsection{Isotropic regime and soliton-like solutions}\label{isotropic}
The simplest way to seek for possible soliton-like solutions associated with the present model on a generic surface is by considering the isotropic regime, $\lambda=0$, and writing down the Hamiltonian (\ref{HamGen}) and its associated equations (\ref{GenForm}) and (\ref{phieq}) in a more suitable form. The assumption of cylindrical symmetry, $\Theta(\zeta,\phi)\equiv\Theta(\zeta)$ and $\Phi(\zeta,\phi)\equiv\Phi(\phi)$, will allow us to get the sine-Gordon equation in a simpler way. In this case, Eq. (\ref{HamGen}) is written as:
\begin{equation}\label{HamIsot}
H_{\text{Isot}}=\int_{-\pi}^{\pi}d\phi\int_{\zeta_1}^{\zeta_2}\left[\left(\partial_{\zeta}\Theta
\right)^2+\sin^2\Theta\left(\partial_\phi\Phi\right)^2\right]d\zeta
\end{equation}
and the Eq. (\ref{phieq}) is simplified to:
\begin{equation}\label{phieq2}
\sin^2\Theta\partial^2_\phi\Phi=0\Rightarrow\Phi=\mathcal{Q}\phi+\phi_0,
\end{equation} 
where $\mathcal{Q}\in \mathbb{Z}$ and $\phi_0$ is a constant of integration. Other possible solutions for the Eq. (\ref{phieq2}) are $\Theta=0$ and $\Theta=\pi$, which are the two vacua of the isotropic Heisenberg model. By taking $\mathcal{Q}=1$ in the solution given by the Eq. (\ref{phieq2}), we can rewrite the Eq. (\ref{GenForm}) as:
\begin{equation}\label{sine-Gordon-Eq}
2\partial_{\zeta}^{2}\Theta=\sin2\Theta,
\end{equation}
which consists in the sine-Gordon equation. By imposing the boundary conditions $\Theta(-\infty)=0$ and $\Theta(\infty)=\pi$, the solution for this equation is a topological soliton excitation, which appears as a continuous transition connecting the two neighbouring minima \cite{rajaraman}:
\begin{equation}\label{solicat}
 \Theta(\zeta)=2\arctan\left(e^{\zeta}\right).
\end{equation}

It is important to note that, usually, the soliton presents a characteristic length scale (CLS) that depends on the characteristic length of the underlying manifold, which, in general, appears in front of the $\sin(2\Theta)$ term in Eq. ({\ref{sine-Gordon-Eq}). However this characteristic length is given by $g_{\phi\phi}/g_{zz}$, which, with our assumptions, is embedded in the $\zeta$ parameter. Thus, the soliton would have its CLS rescheduled to one, e.g., if we take the cylinder case, the CLS of the soliton is equal to the radius $\rho_0$ of the cylinder, however, when working on an infinite cylinder, the change of variable $z\rightarrow z/\rho_0$ eliminates any dependence on $\rho_0$ \cite{cylinder}.

From the calculation of the energy associated to the Eq. (\ref{solicat}), we obtain:
\begin{equation}\label{ensolicat}
 E_{\text{s}}=2\pi J\int_{\zeta_{1}}^{{\zeta_{2}}}\left[(\partial_{\zeta}\Theta)^2+\sin^{2}\Theta\right]d\zeta =\left.-\frac{8\pi J}{e^{2\zeta}+1}\right|_{\zeta_{1}}^{{\zeta_{2}}},
\end{equation}
which is in agreement with the saturated Bogomol’nyi inequality \cite{BogIneq} ($E_{\text{s}}\geq 8\pi J$), once, if we evaluate the solitonic charge, 
\begin{equation}
Q=\frac{1}{4\pi}\int \sin \Theta d\Theta d\Phi,
\end{equation}
in the Eq. (\ref{solicat}), we exactly obtain $Q=-1/(\text{e}^{2\zeta}+1)$.
At the interval $\zeta=(-\infty,\infty)$, the soliton agrees with its counterpart, which lies in an infinite cylinder and represents a complete mapping from the spin sphere to the target manifold, a $\pi$ soliton, so corresponding to the first homotopy class of the second homotopy group of the mapping of the spin sphere. However, when we consider finite surfaces, such a mapping is incomplete and a fractional soliton solution appears, such that no homotopy arguments can be used for classifying solution (\ref{solicat}) as a topological excitation. Indeed, in this case, we must take into account the topology of the geometrical support. In other words, for non-simply connected surfaces, the soliton acquires a finite characteristic length, which prevents its collapse and, consequently, its size does not vanish. Similar scenarios are provided by the annulus, the truncated cone \cite{cone}, the punctured pseudosphere \cite{pseudosphere} and torus \cite{torusmeu}.

Periodic soliton solutions can also be obtained for these rotationally symmetric surfaces \cite{cylinder,Dandoloff-JphysA} and the differences among the particle-like excitations in these and other geometries are associated to their characteristic length scales  \cite{Dandoloff-JphysA}.

Now, we may wonder whether another solitonic solution, with $\Theta(\zeta,\phi)\equiv\Theta(\phi)$ and $\Phi(\zeta,\phi)=\Phi(\zeta)$, should not also appear in this framework. In fact, once $\zeta$ is not, in general, a periodic parameter, $\sin\Phi$ would be periodic only for $\zeta=2\pi m$ (with $m\geq1$ integer) and consequently, the spin sphere mapping should be possible only for this particular case. Besides, the geometry-imposed boundary conditions must lead to a periodic solution for $\Theta$, that is, $\Theta(0)=\Theta(2\pi)$, and solitonic excitations must not occur this way.

\subsection{XY model}\label{discussion}

\begin{figure}
\includegraphics[scale=0.26]{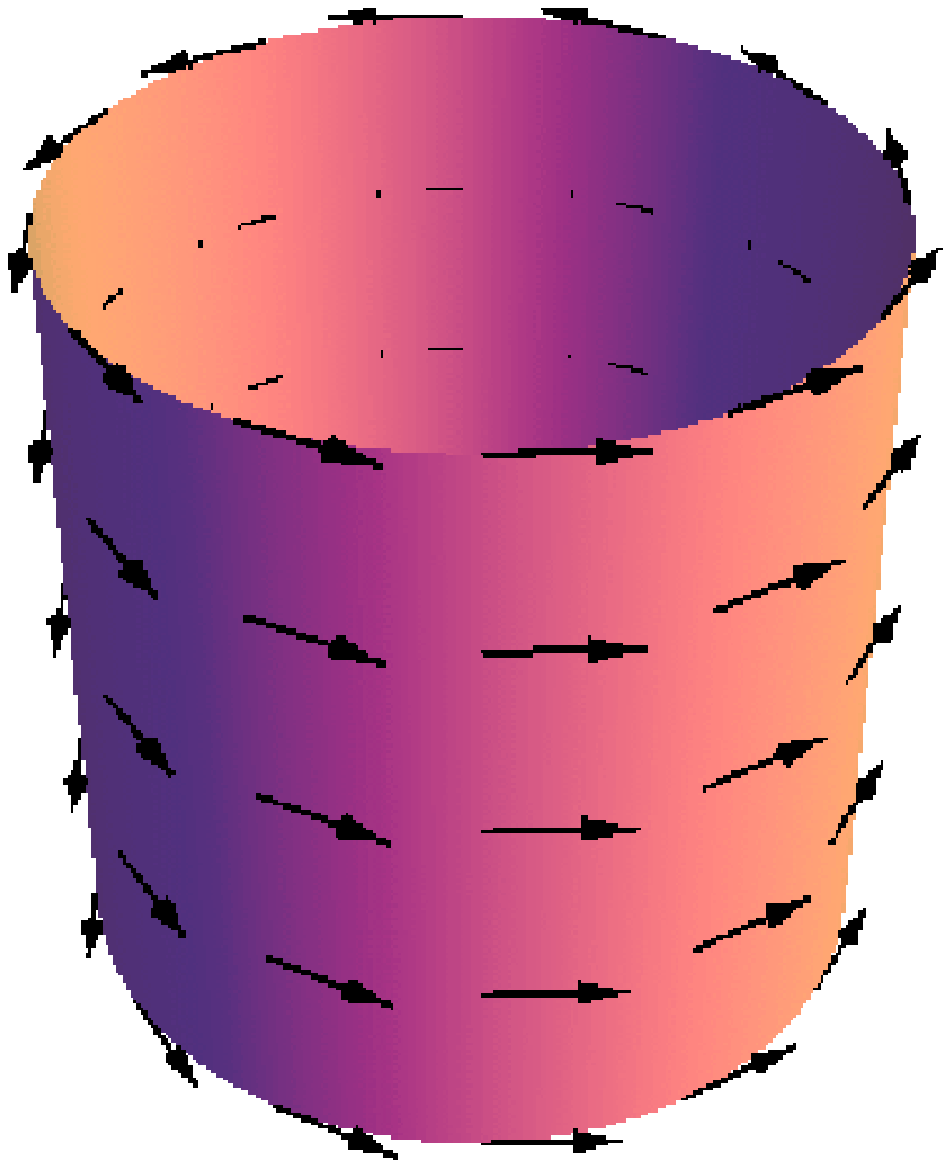}\includegraphics[scale=0.3]{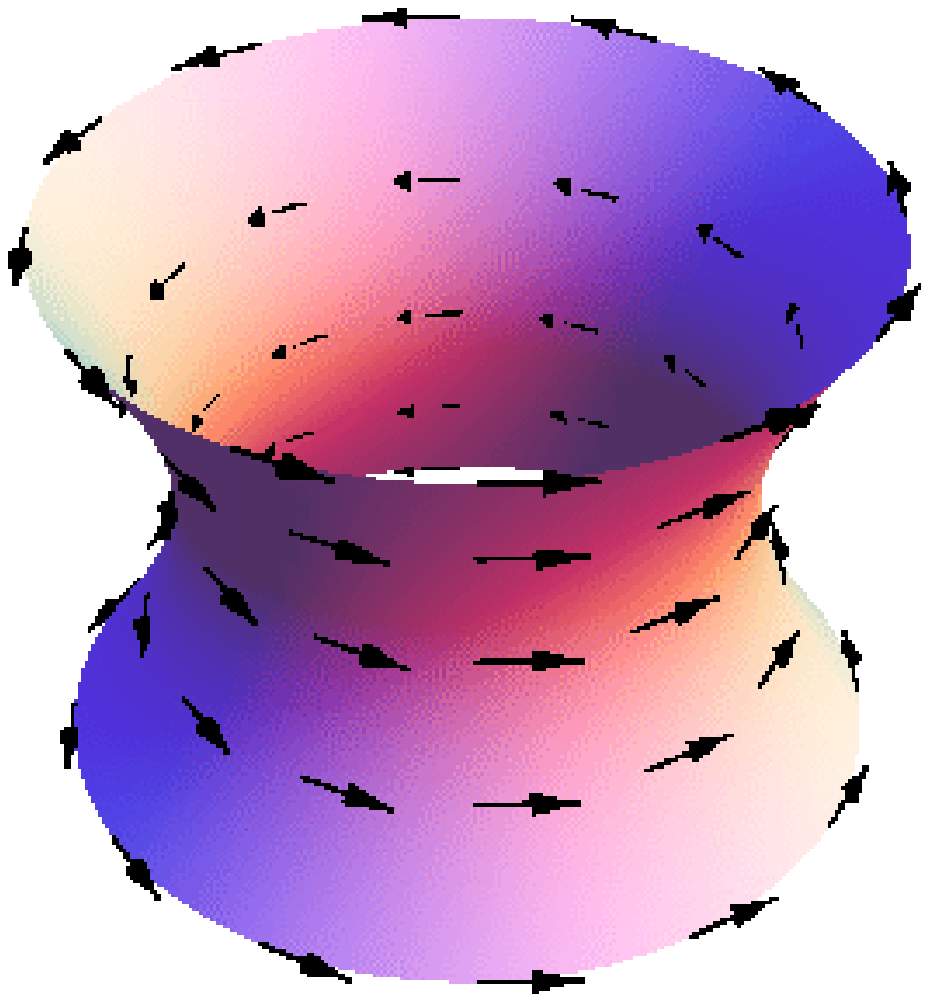}\includegraphics[scale=0.3]{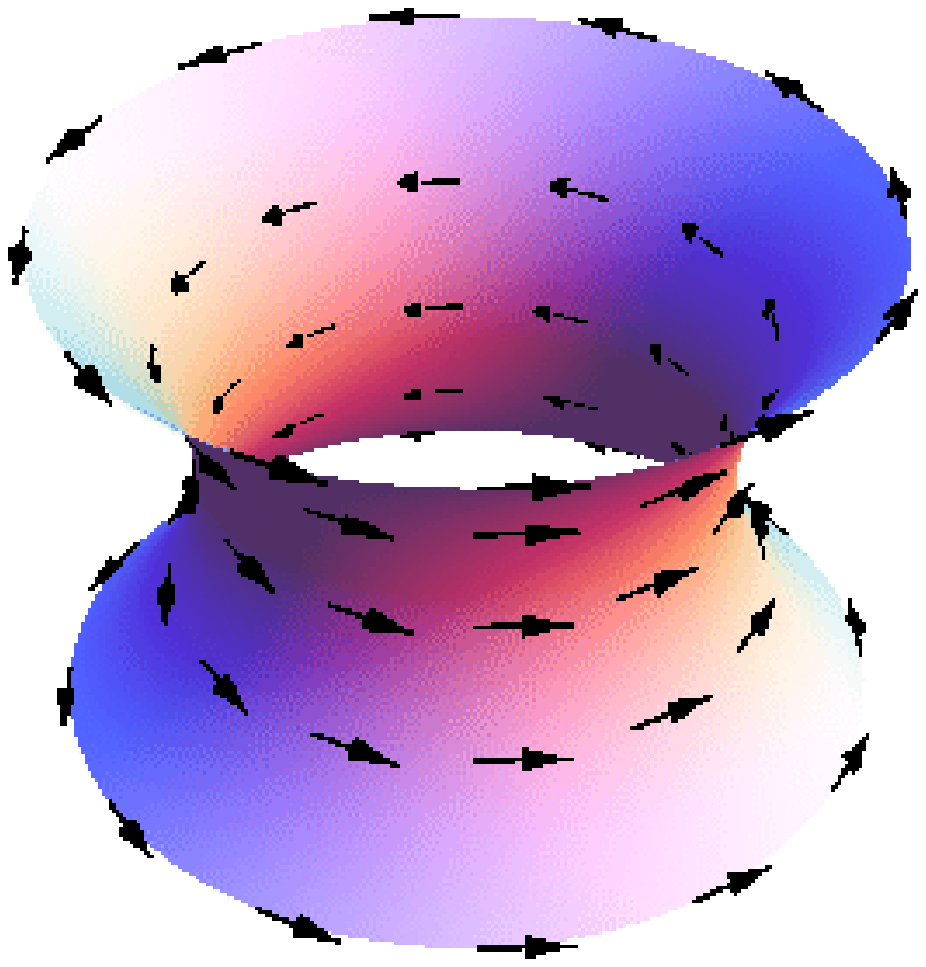}\includegraphics[scale=0.3]{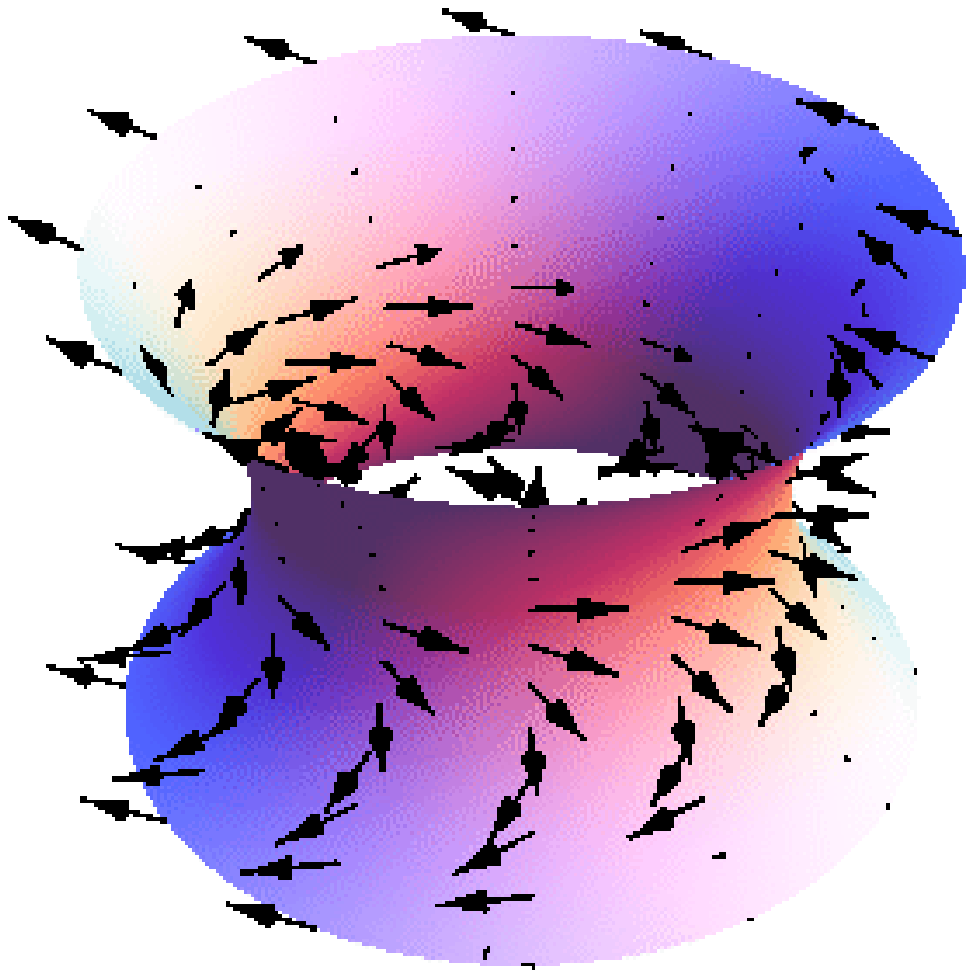}\caption{[Color online] From left to right, the three first figures show the vortex pattern with winding number $\mathcal{Q}=1$ on the surfaces of the cylinder, hyperboloid and catenoid. The last figure represents a helical-like state on the catenoid. In the three first cases, the spins turn around the surface in a closed way. In spite of the cylinder and catenoid have different shape, the characteristic length associated to these are equal, then they present the same value for the vortex energy. On the other hand, the catenoid and hyperboloid surfaces have similar shapes, but they have different characteristic length and still, the mean curvature of the catenoid is null, while the mean curvature of the hyperboloid is nonzero. This leads to different values for the vortex energy on these surfaces. The helical-like state can present its energy equal to the vortex state, provided we have $\Phi(\zeta)=\Phi(-\zeta)$.}\label{VortexPatt}
\end{figure}

Geometrically, a vortex with winding number $\mathcal{Q}$ may be viewed as a set of spins rotating in a closed circuit around a core, whose center is a singular point, or a topological obstruction, which makes it impossible to change the configuration to a perfectly aligned state without tampering with spins at an arbitrary distance from the core. Thus, once a vortex-like configuration cannot be continuously deformed to the ground state, it acquires the status of a topologically stable excitation. 

When we model a vortex as a continuum of spins, we intend to describe only its outer region; once inside the core, the analytical treatment is expected to give only an estimate of its energy, shedding no light about its real structure and spins arrangement, which require numeric or simulation techniques. However, as we shall see in the two particular examples treated here, non-simply connected topologies lead to a natural cutoff for the vortex, which is provided by the surface hole and the solution presents no core. On the other hand, in the case of a connected surface, a singular core takes place, appearing at the surface's self-intercepting points.

It is easy to note that a vortex solution could be obtained from the isotropic Heisenberg Hamiltonian, however, we have chosen to work with the XY model, by adopting $\lambda=-1$ in the adopted model. This choice is justified from the fact that it inserts constraints in the solutions for $\Theta$, which stays confined to the $xy$-plane. In this case, assuming cylindrical symmetry, the Eqs. (\ref{GenForm}) and (\ref{phieq}) are simplified to:
\begin{equation}\label{thetaeq1}
 2\cos^{2}\Theta\partial_{\zeta}^{2}\Theta=\sin2\Theta\left[(\partial_\phi\Phi)^2+
(\partial_{\zeta}\Theta)^{2}-\frac{4}{a^2}\sqrt{\frac{g^{zz}}{g_{\phi\phi}}}\,\right]
\end{equation}
and
\begin{equation}\label{phieq3}
\sin^2\Theta\partial^2_{\phi}\Phi=0,
\end{equation}
and the Hamiltonian (\ref{HamGen}) can be rewritten as:
$$H_{\text{XY}}=J\iint\left[(1-\sin^2\Theta)(\partial_\zeta\Theta)^2+\sin^2\Theta(\partial_\phi
\Phi)^2\right]d\zeta d\phi$$
\begin{equation}\label{hamxy2}
+\frac{4J}{a^2}\iint g^{zz}\cos^2\Theta d\zeta d\phi.
\end{equation}

As previously noted, the simplest solution for the equations (\ref{thetaeq1}) and (\ref{phieq3}) must be given by $\Theta=0$ or $\Theta=\pi$. However, the last term in Hamiltonian (\ref{hamxy2}) gives a positive contribution for the energy, and it is minimized when $\Theta=\pi/2$, which is also a possible solution for (\ref{thetaeq1}). Thus, in order to diminish the energy, the $\Theta$ function must be confined to the $xy$-plane and, in this case, the Eq. (\ref{phieq3}) yields:
\begin{equation}\label{phivortex}
\Phi=\mathcal{Q}\phi+\phi_0, \hspace{1cm}\mathcal{Q}\in\mathbb{Z},
\end{equation}
which represents a vortex with winding number $\mathcal{Q}$. The winding number (vorticity) is formally defined, in the continuum limit, as:
\begin{equation}
\mathcal{Q}=\frac{1}{2\pi}\oint_{C} (\vec{\nabla}\Phi)\cdot d\vec{l},
\end{equation}
where the integration is evaluated along a closed path $C$ around the surface. The vortex pattern with $\mathcal{Q}=1$ for the surfaces of catenoid, hyperboloid, and cylinder can be viewed in the Fig. \ref{VortexPatt}. From Eq. (\ref{hamxy2}), one can see that the vortex energy increases with $\mathcal{Q}^2$, thus, vortices with $\mathcal{Q}>1$ are physically unstable, despite they have topological stability. Thus, we will analyse only the case $\mathcal{Q}=1$.

The vortex energy, for an cylindrically symmetric surface, can be determined from Eq. (\ref{hamxy2}), and it is evaluated to give:
\begin{equation}\label{EnergyVortex}
 E_{\text{vortex}}=2\pi J\zeta\left|_{{_\zeta}_{_1}}^{^\zeta{_2}}\right.,
\end{equation}
where one can note that the vortex energy depends on the $\zeta$ parameter, which is closely related to the curvature of the surface and to the radius of the surface at height z. For example, the torus, sphere and pseudosphere present this parameter given by $\zeta_{\text{tor}}$, $\zeta_{\text{sph}}$ and $\zeta_{\text{psph}}$, given in the set of Eqs. (\ref{ZetaPar}). 

Now, we can look for the possibility of the existence of helical-like states on the considered surfaces, which can be obtained by taking $\Phi(\zeta,\phi)\equiv\Phi(\zeta)$. In this case, the Hamiltonian (\ref{HamGen}) yields:
$$H_{\text{XY}_2}=J\iint\left[(1-\sin^2\Theta)(\partial_\zeta\Theta)^2+\sin^2\Theta(\partial_\zeta
\Phi)^2\right]d\zeta d\phi$$
\begin{equation}\label{hamxy}
+\frac{4 J}{a^2}\iint g^{zz}\cos^2\Theta d\zeta d\phi,
\end{equation}
and the derived Euler-Lagrange equations are:
\begin{equation}\label{thetaeq}
 2\cos^{2}\Theta\partial_{\zeta}^{2}\Theta=\sin2\Theta\left[(\partial_\zeta\Phi)^2+
(\partial_{\zeta}\Theta)^{2}-\frac{4g^{zz}}{a^2}\,\right].
\end{equation}
Using the same arguments of the previous analysis, it is easy to note that the simplest solution minimizing the energy is given by $\Theta=\pi/2$. In this way, the Eq. (\ref{phieq}) is simplified to:
\begin{equation}\label{phieq4}
\partial^2_{\zeta}\Phi=0\rightarrow\Phi(\zeta)=q\zeta+\zeta_0,
\end{equation}
where one can note that this solution looks like that given in Eq. (\ref{phivortex}). However, unlike the vortex case, $q$ must not be an integer number, once $\zeta$ is not, in general, periodic. Thus, we have not a closed path along the surface to define the topological charge. In order to obtain a helical-like state rotating by $2\pi$ along the surface, we can assume the boundary condition $\Phi(\zeta)=\Phi(-\zeta)$. Thus, $q$ would be an integer and the spin vector field form a closed structure around the $xy$-plane of the internal space when the surface is mapped from $-z$ to $z$ (See Fig. \ref{VortexPatt}). The energy of this spin configuration, for $q=1$, is given by:
\begin{equation}\label{HeliState}
E_{\text{Hel}}=2\pi J\zeta,
\end{equation}
which is the same solution found for the vortex state. It is important to note that the helical-like state has not topological stability, once $q$ must not be considered as a topological charge. The coincidence in the energy values of the vortex and helical-like states appears due the imposed boundary conditions for $\Phi(\zeta)$. Another important fact is that, if we were considering the magnetostatic energy term in our model, this configuration would have magnetic energy greater than that of the vortex state due the appearing of surface magnetic charges, $\sigma=\mathbf{m}\cdot\mathbf{n}$, and it would not be stable.

Finally, once the Eq. (\ref{phieq3}) gives us two possible solutions to $\Phi(\phi)$, we can look for the critical value for the anisotropy parameter, $\lambda_c$, such that for $\lambda<\lambda_c$ one has solution (\ref{phivortex}) and for $\lambda > \lambda_c$ the
solution $\Theta=0$ appears. Here, $\lambda_c$ will be determined by calculating the energy for these different spin configurations and comparing them. The stable state will be that one which minimizes the energy. With this assumptions, it is easy to note that \begin{equation}\lambda'_c=-{\zeta}/{\xi},\end{equation}
where $\xi=\int\sqrt{g^{zz}/g_{\phi\phi}}dz$ and $\lambda'_c=4\lambda_c/a^2$. Thus, it can be noted that, as well as the vortex energy, $\lambda_c$ also depends on the surface's geometrical properties and, consequently, different surfaces must present different values for $\lambda_c$. In the case of the cylinder it is easily shown that $\lambda'_c=-1$. 

In order to present some examples showing the relation among curvature, vortex energy and critical anisotropy parameter, we will study two particular surfaces: the catenoid and the hyperboloid, which are non-simply connected and negatively curved surfaces, having cylindrical symmetry.

\section{Two particular cases}\label{partcases}
\subsection{Catenoid}
The first surface to be considered is the catenoid, which is a nonplanar minimal surface. The catenoid has mean curvature everywhere zero and has the fascinating property that it can be deformed into a helicoid in such a way that every surface along the way is a minimal surface, which is locally isometric to the helicoid. This surface appears in many physical systems, e.g., a soap film formed between two coaxial rings takes on this shape \cite{Soap-PRE} and in membrane fissions, where the characteristic of the neck shape is determined by the relationship between the neck radius and the monolayer thickness \cite{Koslovski-BJ}. 

In order to describe the adopted model in the geometry of the catenoid, we have parametrized this surface as follows:
\begin{equation}
 x=\rho_{_0}\cosh\left(\frac{z}{\rho_{_0}}\right)\cos\phi,\hspace{1cm}y=\rho_{_0}\cosh\left(\frac{z}{\rho_{_0}}\right)\sin\phi,
\end{equation}
where $\phi\in[0,2\pi]$, $z\in(-\infty,\infty)$ and $\rho_{_0}$ is the radius of the circle in the $z=0$ plane. This parametrization yields the negative Gaussian curvature $G_{\text{cat}}=-{1}/{\rho_{_0}^2}\text{sech}{^4}({z}/{\rho_{_0}})$ and it allows us to get the characteristic length scales of the catenoid, which are given by:
\begin{equation}\label{zetaCat}
\zeta_{\text{cat}}=\int_{-h}^{h}{\frac{1}{\rho_{_0}}dz}=\frac{2h}{\rho_{_0}},\hspace{1cm}\text{and}\hspace{1cm}\xi_{\text{cat}}=2\coth\frac{h}{\rho_{_0}},
\end{equation}
where $2h$ is the height of the catenoid. From the Eq. (\ref{zetaCat}), one can see that the critical value for the anisotropy parameter, in the catenoid case, depends on $h$, and it is given by: 
\begin{equation}
\lambda'\,^{\text{cat}}_c=-\frac{h}{\rho_{_0}}\coth\left(\frac{h}{\rho_{_0}}\right).
\end{equation}

It can be noted that, unlike the cylinder case, the critical anisotropy parameter that stabilizes the vortex configuration depends on the relation between height and central radius (CR) of the catenoid, represented by $\rho_{_0}$. For $h\rightarrow0$, we get $\lambda'\,^{\text{cat}}_c\rightarrow-1$ and, for large values of $h/\rho_{_0}$, it can be noted that $\lambda'\,^{\text{cat}}_c$ decreases linearly with the height. This behaviour is expected, once the vortex energy, on this surface, is a linear function of $h$ (See Eq. (\ref{vortex-energy-cat})). The Fig. \ref{lambdabehaviour} show the behaviour $\lambda'_{c}$ with when we varies the height and CR of the catenoid surface. One can note that $\lambda'\,^{\text{cat}}_c\rightarrow-1$ for large $\rho_{_0}$ values, as expected, once, when $\rho_{_0}\gg h$, the catenoid surface looks like a cylinder.
\begin{figure}
\includegraphics[scale=0.2]{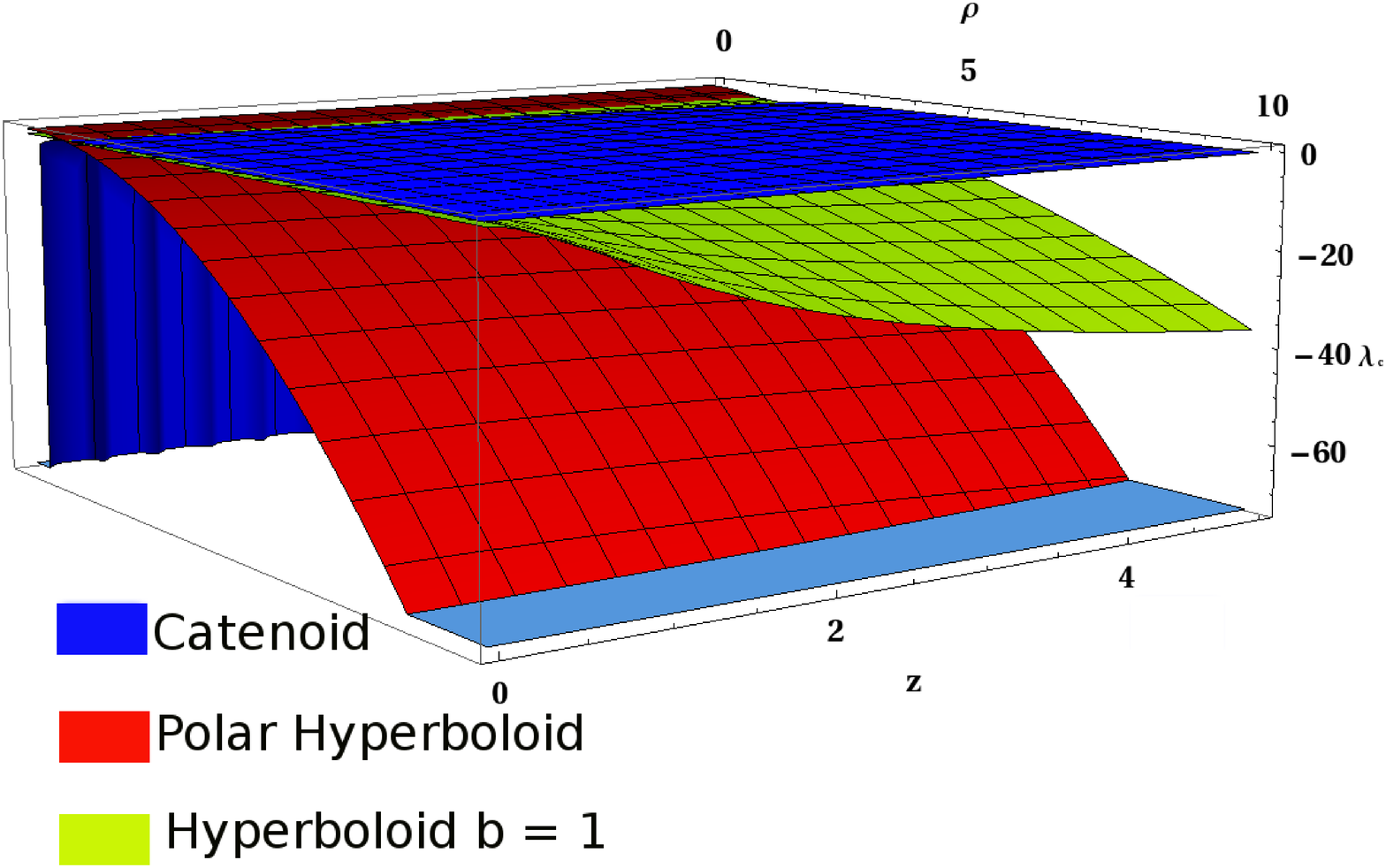}\includegraphics[scale=0.2]{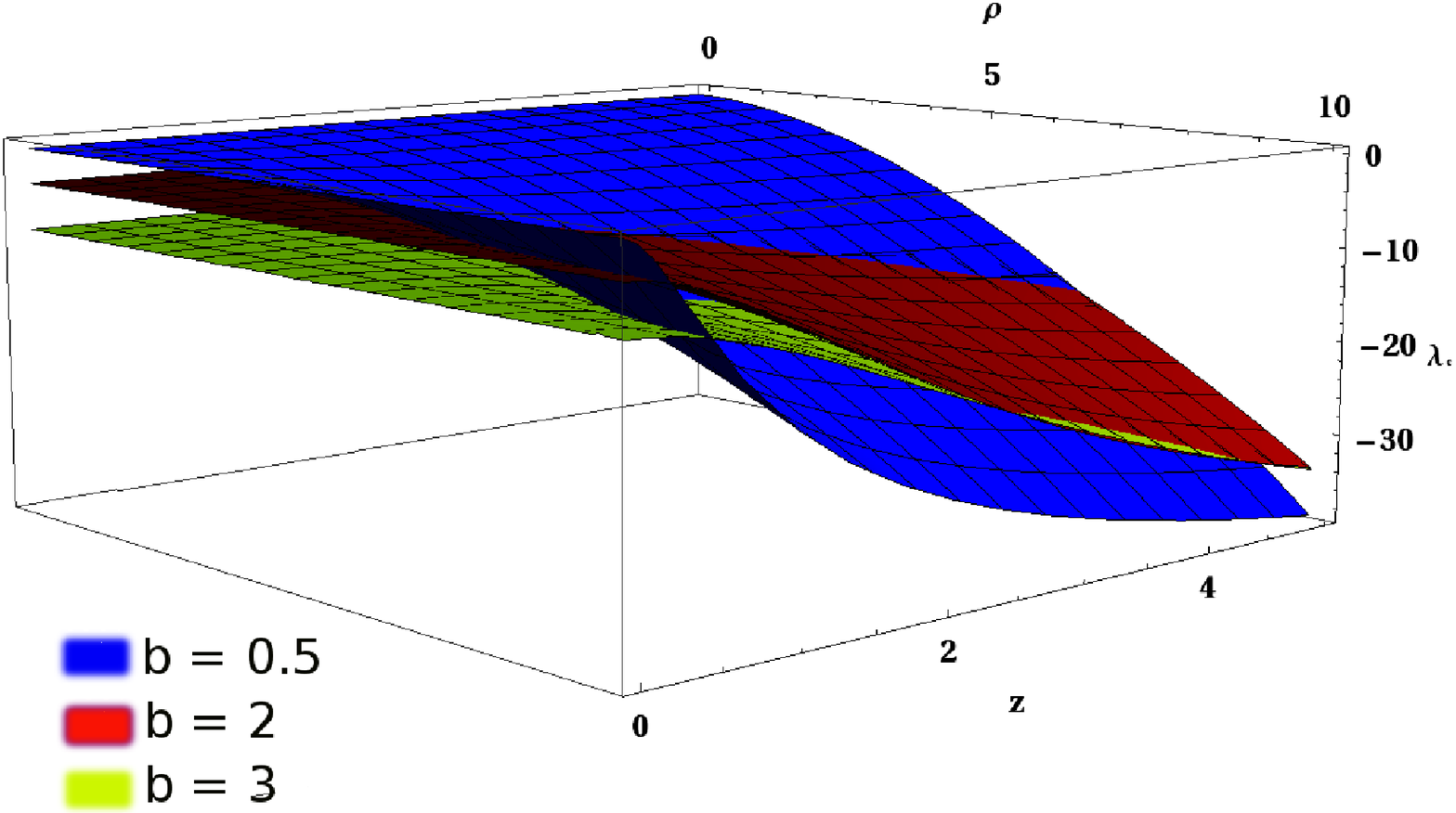}\\\includegraphics[scale=0.4]{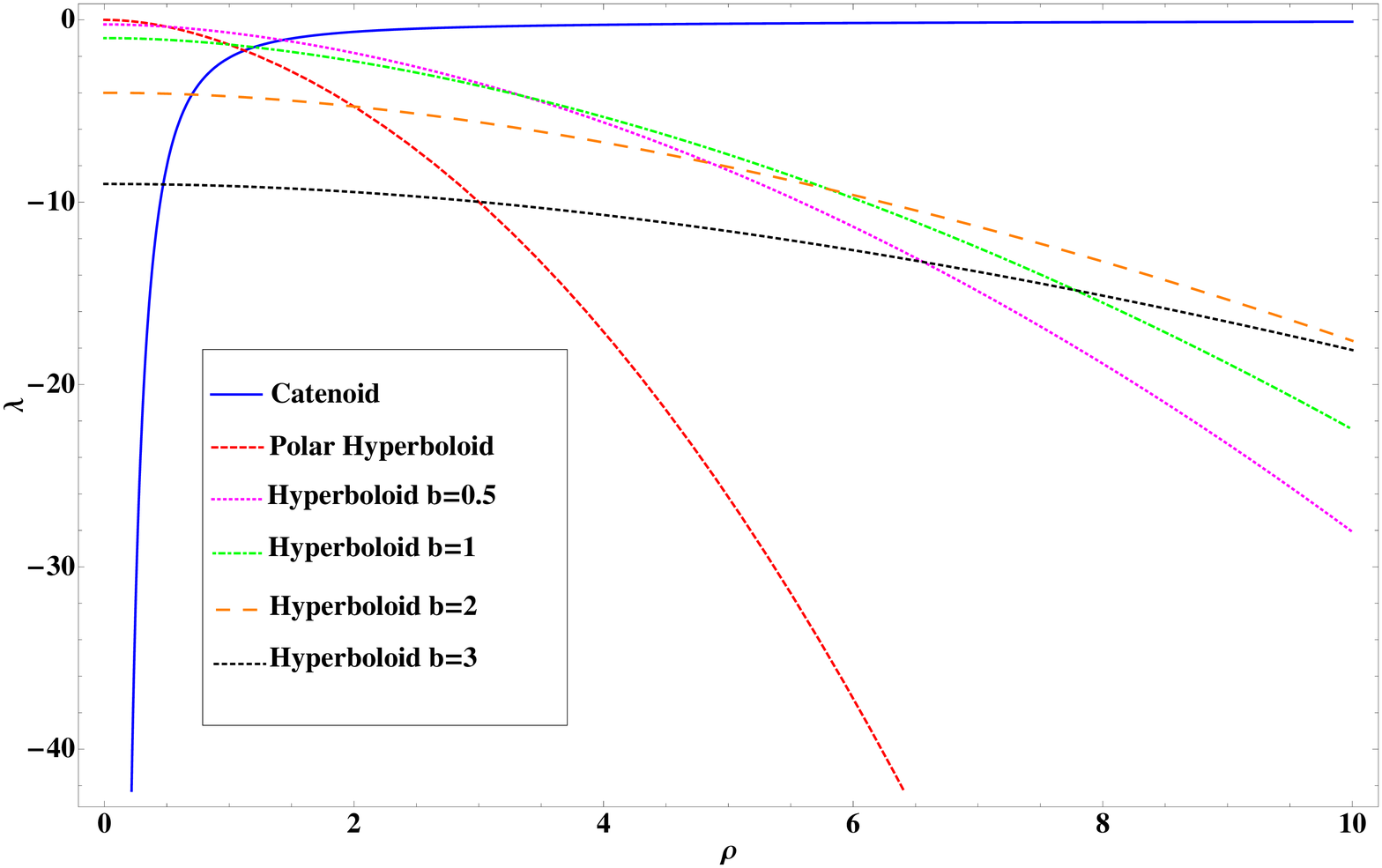}\caption{[Color online] The top left figure show the critical value for the anisotropy parameter, in function of height and CR for the catenoid, polar hyperboloid and hyperboloid with $b=1$. Note that the catenoid and polar hyperboloid present $\lambda'_c$ strongly depending on $\rho_{_0}$. The top right figure show $\lambda'_c$ for the hyperboloid with $b=0.5$, $b=2$ and $b=3$. In the bottom figure, we have fixed the height $z=4$ and CR varies from $0$ to 10. One can see that critical anisotropy parameter is highly depending on the geometrical properties of the surface. Furthermore, it increases with the radius, in the case of the catenoid, it decreases for the hyperboloidal surfaces. If we consider different heights, there was no qualitative changes in the $\lambda'_c$ behaviour.}\label{lambdabehaviour}
\end{figure}

\subsection{Hyperboloid}

The hyperboloid is a quadratic surface that may be one- or two-sheeted. The one-sheeted hyperboloid is a surface of revolution obtained by rotating a hyperbola about the perpendicular bisector to the line between the foci, that is, about the $z$ axis, while the two-sheeted hyperboloid is a surface of revolution obtained by rotating a hyperbola about the line joining the foci. 

When oriented along the $z$-axis, the one-sheeted circular hyperboloid with skirt radius $\rho$ has its parametrization given by:
\begin{equation}\label{hyppar}
 x=\rho\sqrt{1+\eta^{2}}\cos\phi,\hspace{1cm}
y=\rho\sqrt{1+\eta^{2}}\sin\phi\hspace{1cm}\text{and}\hspace{1cm}z=b\eta.
\end{equation}

This parametrization yields a shape similar to the catenoid, however the last one has mean curvature null everywhere, while that one of the hyperboloid is given by \cite{Wolfram}:
\begin{equation}
K_{\text{hyp}}=\frac{b^2[\rho^2(\eta^2-1)+b^2(1+\eta^2)]}{2\rho[\rho^2+b^2(1+\eta^2)]^{(3/2)}},
\end{equation}
and the Gaussian curvature is:
\begin{equation}
G_{\text{hyp}}=-\frac{b^2}{b^2+(b^2+\rho^2)\eta^2}.
\end{equation}
Thus, unlike the catenoid, the hyperboloid is not a minimal surface. Furthermore, there are substantial differences in their $(x,y)$ coordinates away from the plane $z=0$, as well as in their geometrical properties.

The general parametrization given in Eq. (\ref{hyppar}) leads to a hard integral to be calculated, so that, when necessary, $\zeta_{\text{hyp}}$ and $\xi_{\text{hyp}}$ will be computed numerically. However, a particular and interesting kind of hyperboloid, which leads to solvable integrals to the characteristic length parameters, is described by the polar hyperbolic coordinate system (biharmonic coordinates), which can be obtained for taking $b=\rho$ in the Eq. (\ref{hyppar}). This particular coordinate system was recently used to develop the pseudospherical functions on an one-sheeted polar hyperboloid \cite{Kowalski-JPA}. These functions may be important if we consider problems where the solution of the Laplace equation in systems with hyperbolic symmetry is demanded. For instance, they could be used to calculate the magnetostatic energy in magnets with hyperbolic shape or still, to calculate the magnetic field inside solenoids with this geometry, when it is traversed by an electric current and/or still, to determinate the electric field generated by a charged polar hyperbolic shell. 

By using the parametric equations (\ref{hyppar}), the vortex energy to the polar hyperboloid is given by the Eq. (\ref{EnergyVortex}) with $\zeta$ given by:
\begin{equation}\label{zetaHyp}
\zeta_{\text{phyp}}=\int{\chi(\eta)d\eta}=\sqrt{2}\text{arcsinh}({\sqrt{2} \eta}) + 
   \frac{1}{4}\ln\left(\frac{1 + 3 \eta^2 - 2 \eta \sqrt{1 + 2 \eta^2}}{1 + 3 \eta^2 + 2 \eta \sqrt{1 + 2 \eta^2}}\right)+\kappa_2,
   \end{equation}
where \begin{equation}\label{chi}
\chi(\eta)=\frac{\sqrt{1+2\eta^2}}{1+\eta^2}.
\end{equation}
and $\kappa_2$ is a constant of integration. 

Finally, we must calculate the critical anisotropy parameter of this surface. In order to do this, we have gotten $\xi_{\text{hyp}}$ numerically. The results of the critical anisotropy parameter for the polar hyperboloid, and the hyperboloid with $b=0.5$, $b=1$, $b=2$ and $b=3$ are summarized in the Fig. \ref{lambdabehaviour}. As well as the catenoid, $\lambda'\,^{\text{hyp}}_c$ depends on the height and radius of the surface. However, the anisotropy parameter must decreases in order to give stability to the vortex state. Note that for small CR, $|\lambda'\,^{\text{hyp}}_c|$ has its smallest value for $b=0.5$, when compared to the cases where $b=1$, $b=2$ or $b=3$. However, when we increases the $\rho_{_0}$ value, this situation is reversed. Finally, the analysis of the Fig. \ref{lambdabehaviour}\ \ show that $\lambda'\,^{\text{hyp}}_c\rightarrow-1$ only for $h\rightarrow0$ and $\rho_{_0}\rightarrow0$.

%===============================================================================================
%===============================================================================================
%===============================================================================================

\subsection{Comparison between the vortex energy on a catenoid and hyperboloid}

From the Eq. (\ref{EnergyVortex}) and $\zeta_{\text{cat}}$, we get the energy of a vortex on a catenoid surface, which can be explicitly written as:
\begin{equation}\label{vortex-energy-cat}
E_{\text{vortex}}^{\text{cat}}=\frac{4\pi J h}{\rho_{_0}},
\end{equation}
which is the same energy obtained for a vortex on the surface of a cylinder with height $2h$ and radius $\rho_{_0}$. This is an interesting result, since the area of the catenoid is greater than that of the cylinder. This fact can be explained because these surfaces present the same characteristic length $1/\rho_{_0}$, which can be noted by the development of the Hamiltonian (\ref{HamGen}) for the cylinder surface. Furthermore, from the analysis of the Fig. \ref{VortexPatt}, it can be noted that, in the $z=0$ plane, the neighbour spins on the catenoid must have the angle equal to that of the spins on the cylinder, however, when $z\neq0$, the neighbour spins that turning around the catenoid surface have a lower deviation one to another, when compared with that of the cylinder, diminishing the exchange energy (given by the Heisenberg model) in this plane, which compensates the largest area. To say this on another way, if the circle situated in the plane $z=h$, for each surface, is divided in points with a distance $\Delta x$ one to another, we have that the circle on the cylinder is divided in $n'=2\pi \rho_{_0}/\Delta x$ points, while the circle on the catenoid is divided in $n=2\pi\rho_{_0}\cosh(h/\rho_{_0})/\Delta x$. In this way, we have that $n=n'\cosh(h/\rho_{_0})$. On the other hand, the obtained circumferences can be divided in arcs of angles $\theta=2\pi/n$ and $\theta'=2\pi/n'$ for the catenoid and cylinder, respectively. It is immediate to note that $\theta=\theta'/\cosh(h/\rho_{_0})$. A closed spin texture on these circles can be given by associating one spin for each point, in such way that the angle between two neighbour spins for the catenoid (cylinder) is $\theta$ ($\theta'$). The energy for these vortex-like structure in the circle on the catenoid is $E_{\text{cat}}=\sum_{i,j}\cos\theta_{ij}=n\cos\theta$, where the subscripts $i$ and $j$ indicate neighbour spins. The energy for the circle on the cylinder is $E_{\text{cyl}}=n'\cos\theta'$. Finally, $E_{\text{cat}}$ and $E_{\text{cyl}}$ are related by:
\begin{equation}
E_{\text{cat}}=\frac{n\cos\frac{2\pi}{n}}{n'\cos\frac{2\pi}{n'}}E_{\text{cyl}}.
\end{equation}
For $n\gg 1$, we have that $E_{\text{cat}}=E_{\text{cyl}}\cosh(h/\rho_{_0})$. Now, the cylinder and catenoid heights can be divided in $m'$ and $m$ circles, respectively. The total energy of the vortex on the surfaces are given by:  
\begin{equation}
E_\text{tcyl}=m'E_{\text{cyl}}\hspace{1cm}\text{and}\hspace{1cm}E_\text{tcat}=mE_{\text{cat}}=mE_{\text{cyl}}\cosh\frac{h}{\rho_{_0}}.
\end{equation}
Since $m=m'\frac{\rho_{_0}}{h}\sinh\frac{h}{\rho_{_0}}$, we obtain:
\begin{equation}
E_\text{tcat}=\left(\frac{\rho_{_0}}{2h}\sinh\frac{2h}{\rho_{_0}}\right)E_{\text{tcyl}}.
\end{equation}
If $\frac{2h}{\rho_{_0}}\ll 1\rightarrow\sinh\frac{2h}{\rho_{_0}}\approx\frac{2h}{\rho_{_0}}$, and finally $E_{\text{tcat}}=E_{\text{tcyl}}$, as we wanted to show. This discrete analysis is not valid for large $2h/\rho_{_0}$ values, once in this limit, the approximation of the surfaces by points does not represent the continuum approach to the Heisenberg Hamiltonian and the gotten results must not agree with that found in the Eq. (\ref{vortex-energy-cat}).

In the case of the hyperboloid, analytical calculations will be done only for the biharmonic coordinates. In this case, in order to obtain and analyse the vortex energy on a hyperboloid with height $2h$, we will take the limits $\zeta_1=-h/\rho_{_0}$ and $\zeta_2=h/\rho_{_0}$ in the Hamiltonian (\ref{hamxy}). Thus, the vortex energy is given by:
\begin{equation}\label{VortEnergyphyp}
E_{\text{vortex}}^{\text{phyp}}=4\pi J\left[\sqrt{2}\text{arcsinh}\left(\sqrt{2}\frac{h}{\rho_{_0}}\right)+\frac{1}{4}\ln\left(\frac{1+3\left(\frac{h}{\rho_{_0}}\right)^2-2\frac{h}{\rho_{_0}}\sqrt{1+2\left(\frac{h}{\rho_{_0}}\right)^2}}{1+3\left(\frac{h}{\rho_{_0}}\right)^2+2\frac{h}{\rho_{_0}}\sqrt{1+2\left(\frac{h}{\rho_{_0}}\right)^2}}\right)\right].
\end{equation}

Despite of the polar hyperbolic coordinates give us an analytical solution to the vortex energy, it also will be interesting to calculate it numerically for the most general hyperboloid ($b\neq\rho_{_0}$) and analyse its behaviour in function of the radius $\rho_{_0}$. In this case, $\zeta_{\text{hyp}}=\int{\chi'(\eta)d\eta}$, where 
\begin{equation}
\chi'(\eta)=\frac{\sqrt{b^2(1+\eta^2)+\rho_{_0}^2\eta^2}}{\rho_{_0}(1+\eta^2)}.
\end{equation} 
The numerical integration was done for using the Simpsons rule with a Fortran code, and the results can be viewed in Fig. \ref{VortexEnergy}.

\begin{figure}
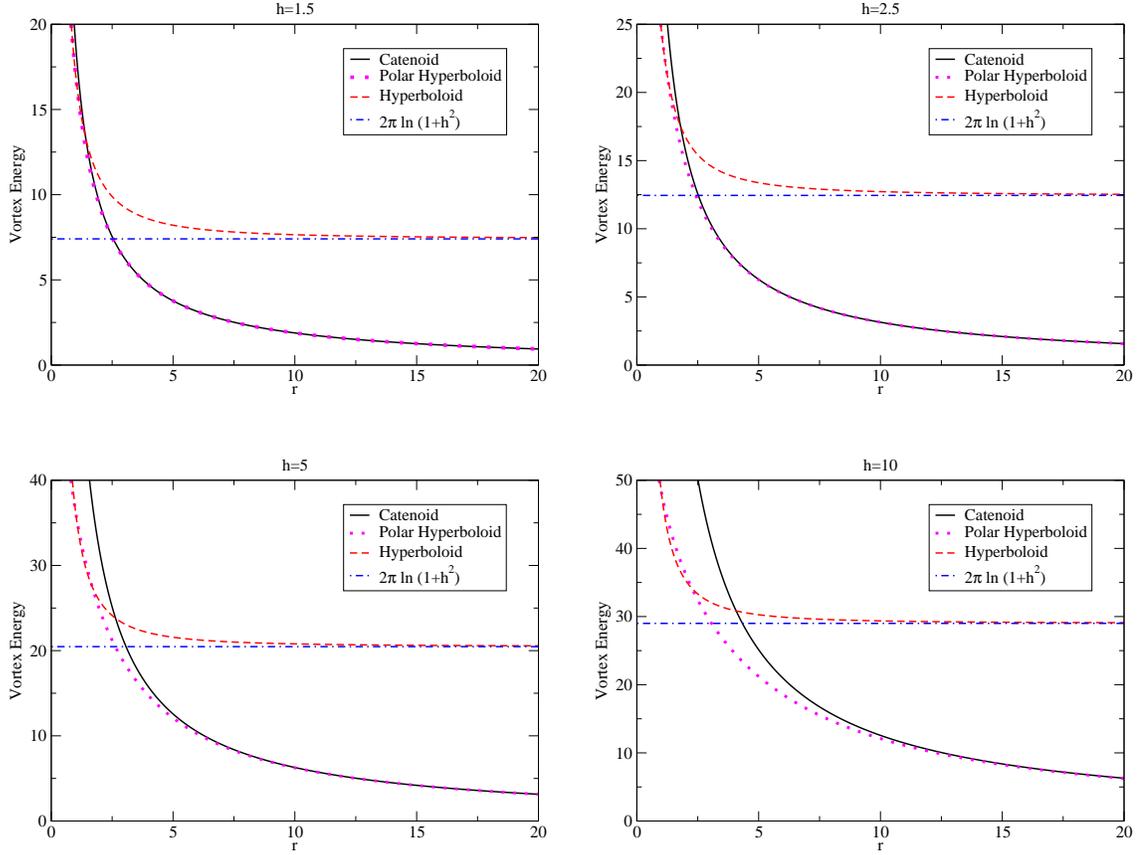
\vspace{0.5cm}
\includegraphics[scale=0.3]{h3.eps}\hspace{1.5em}\includegraphics[scale=0.3]{h5.eps}\vspace{0.8cm}
\includegraphics[scale=0.3]{h10.eps}\hspace{1.5em}\includegraphics[scale=0.3]
{h20.eps}\caption{[Color online] Energy of the vortex on the catenoid (black line) and the hyperboloid (red dashed line) surfaces. Here, we have done $J=1$, $b=1$ and $\rho_{_0}\equiv\rho$ is evaluated in the interval from 0.01 to 20. Four values of $h$ are considered: $h=1.5$ (top left), $h=2.5$ (top right), $h=5$ (bottom left) and $h=10$ (bottom right). One can see that, for small values of CR (central radius), the energy of a vortex on the hyperboloid is lower than that of the vortex on a catenoid. However, for large values of CR, the vortex energy on a catenoid vanishes, while that one of the hyperboloid tends to $2\pi\ln(1+h^2)$. The energy of a vortex on the polar hyperboloid (magenta dots) is always lower than that on a catenoid and, unlike the general hyperboloid ($b\neq\rho$), it vanishes when $\rho\rightarrow\infty$.}\label{VortexEnergy}
\end{figure}

To continue our analysis, it will be useful to define the upper radius (UR), which is the value of the radius of the surfaces in the plane $z=\pm h$. The cylinder, the catenoid and the hyperboloid have their CR and UR related by:
\begin{equation}\label{rel-UPCENT-radii}
R_{_\text{Upper}}^{\text{cyl}}=\rho,\hspace{1cm}R_{_\text{Upper}}^{\text{cat}}=\rho\cosh\left(\frac{h}{\rho}\right)\hspace{1cm}\text{and}\hspace{1cm}
R_{_\text{Upper}}^{\text{hyp}}=\rho\sqrt{1+h^2},
\end{equation}
where $\rho$ represents the CR of the cylinder, catenoid and hyperboloid ($b=1$), respectively.

From the Fig. \ref{VortexEnergy}, one can note that, for small values of $\rho$, the vortex on a hyperboloid has lower energy than that on a catenoid. However, when $\rho$ increases, this behaviour changes, and the energy of a vortex on the hyperboloid becomes greater than that on the catenoid. The value of $\rho$ for the transition point in which the vortex energy on the hyperboloid and on the catenoid intersect themselves depends on $h$, and varies from $\rho\approx2$ for $h=1.5$ to $\rho\approx4$ for $h=10$. This fact can be explained from the differences of the properties of the geometries of the catenoid and hyperboloid when $\rho\rightarrow\infty$ and the height $h$ is maintained constant. While the catenoid has its UR related with the height by $\cosh(h/\rho)$, the UR of the hyperboloid increases with $\sqrt{1+h^2}$, which implies that, for $\rho\gg h$, we have $R^{\text{cat}}_{\text{Upper}}\rightarrow R^{\text{cyl}}_{\text{Upper}}$ and $R^{\text{hyp}}_{\text{Upper}}\rightarrow \infty$. Indeed, for $\rho\gg h$, the catenoid surface looks like a cylinder, while the hyperboloid has the topology and geometry of a plane with a hole. The analysis of the Fig. \ref{VortexEnergy} shows also that  $\rho\rightarrow\infty$  when $E^{\text{hyp}}_{\text{vortex}}\rightarrow2\pi\ln(1+h^2)$. This is the second time that an hyperbolic surface presents an asymptotic finite and nonzero energy when the radius of the surface tends to infinity \cite{pseudosphere}. This finite value to the vortex energy for $\rho\rightarrow\infty$ can be explained because, unlike the catenoid case, the characteristic length of the hyperboloid does not tend to zero in this limit. In fact:
\begin{equation}
\chi'(\eta)_{\rho\rightarrow\infty}=\frac{\eta}{1+\eta^2}\Longrightarrow\zeta_{hyp}^{\rho\rightarrow\infty}=\ln(1+\eta^2).
\end{equation}
The case where $b\neq1$ does not lead to qualitative changes on the vortex energy behaviour on a hyperboloid. In this case, the curve that characterizes the vortex energy is moved upwardly, however, as well as the $b=1$ case, it tends to $2\pi\ln(1+h^2)$, which is associated to the fact that the $b$ value does not affect the asymptotic behaviour of $\zeta_{\text{hyp}}$ when $\rho\rightarrow\infty$ (see Eq. (\ref{chi})).

Finally, we will analyse the vortex energy on the polar hyperboloid surface, which is given by the Eq. (\ref{VortEnergyphyp}). From the Fig. \ref{VortexEnergy}, one can note that this surface presents a vortex whose energy is always lower than that on the catenoid. For small CR, the vortex energy calculated by using biharmonic coordinates is approximately that one obtained for general hyperboloid, once $\rho\approx b=1$. However, as well as the result found for the catenoid, we have that $E_{\text{phyp}}\rightarrow0$ when $\rho\rightarrow\infty$. This fact can be explained because in this limit, the polar hyperboloid looks like a catenoid and its characteristic length is given by: 
\begin{equation}
\chi'(\eta)_{\rho\rightarrow\infty}=1\Longrightarrow \zeta_{\text{phyp}}^{\rho\rightarrow\infty}=z/\rho=\zeta_{\text{cat}}.
\end{equation} 

It is important to note that a vortex, understood as a topological defect, is a singularity in the order parameter. Then, at first sight, it could mean that the obtained solutions, which are smooth spin textures, are not vortices. However, despite we do not have analysed this case explicitly, when $\text{CR}\rightarrow0$, for both surfaces, one get that the vortex energy diverges due a singularity in the order parameter at the center of the surfaces, which is the necessary condition to call these excitations of topological vortices. A way to control this divergence would be to insert a cutoff of length $\ell_{_0}$ at the center of the surfaces, as well as it has been done for other geometries where this singularity appeared \cite{torusmeu,pseudosphere,sphere}. However, unlike the spherical, pseudopherical or toroidal cases, if we introduce this cutoff in the point where the surfaces of the catenoid and hyperboloid self intersecting, these will be divided in two half-surfaces and this case is outside the scope of this work.

If we consider the case where the UR of the three surfaces are equal, it is immediate to note that the catenoid has smaller energy than that of the cylinder, since the vortex energy for both surfaces is proportional to $1/\rho$. However, there are not qualitative changes when we study the hyperboloid case. Here, as well as the previous analysis, the vortex energy on the three surfaces diminishes with the increasing of UR, tending to zero when $r\rightarrow\infty$ (cylinder, catenoid and polar hyperboloid). 

In conclusion, the lowest value found for the vortex energy, for any CR, occurs for the polar hyperboloid, which can indicate, at first sight, that among the geometries considered here, a ``hyperboloidal nanoshell'' could support a vortex-like magnetization with more stability than their cylindrical counterparts. However, to ensure this statement, one must calculate the magnetostatic energy for other possible magnetization states and consider the volume of the magnet.

Once miniaturization is an important issue in nanotechnology, the obtained results may guide future researches on the stability of vortices in circular nanomagnets. About this theme, it has been shown that, due the out-of-plane component, the vortex energy in circular nanodots is greater than that obtained for nanorings with same dimensions, because they do not present the out-of-plane component in the vortex core \cite{Bellegia-JMMM}. In this way, nanorings can support vortex as magnetization ground state for smaller radius than that of nanodots. Furthermore, it has been shown that the smooth curvature of a toroidal device must improve the vortex stability, when compared with cylindrical nanorings \cite{Vagson-JAP}. Thus, it may be possible that, by introducing curvature in the nanorings fabrication in order to obtain ``hyperboloidal nanorings'', the vortex gets more stability, and one could fabricate smaller nanomagnets with a vortex as the ground state.

%===============================================================================================
%===============================================================================================
%===============================================================================================

\section{Conclusions and Prospects}\label{conclusion}

We have studied the anisotropic Heisenberg model on curved surfaces with cylindrical symmetry in order to obtain a class of topological spin excitations. By taking $\lambda=0$, soliton-like solution has been gotten and it have been shown that they have energy obeying the Bogomolnyi's inequality. Fractional solitons, which has not topological stability, have also been predicted to appear if we consider finite surfaces. By taking $\lambda=-1$, one obtain the XY model, whose simplest solution is a vortex with winding number $\mathcal{Q}=1$. The energy and stability associated to vortices is closely linked to the geometric properties of the surface. Furthermore, we have gotten the anisotropy parameter, $\lambda_c$, for which the vortex appears as the groundstate, in comparison with the state in which the spins point in the $z$-axis direction. It has been shown that $\lambda_c$ is also related to the geometry of the substrate and it must be variable for surfaces with non-constant Gaussian curvature. In addition, we have also studied the possibility of the appearing of helical-like states on the considered surfaces. The energy of these states have been calculated and, for periodic boundary conditions, the same energy predict for the vortices has been obtained for this spin configuration, however, helical states does not present a topological charge and homotopy arguments can not be used to ensure its stability. 

In order to exemplify our results, two surfaces were explicitly analysed: the catenoid and the hyperboloid, which are non-simply connected manifolds having negative and variable Gaussian curvature. For each surface, the vortex energy has been compared with that presented for a vortex on the cylinder, which is a geometry that have received large attention in nanomagnetism researches. It has been noted that, for small CR, the general hyperboloid ($b\neq\rho$) presents a vortex energy lower than that presented by the vortex on a cylinder and on a catenoid, however, for $\rho\rightarrow\infty$, unlike the other studied geometries, the vortex energy on a hyperboloid does not vanish. Furthermore, the smallest value for the vortex energy is given by the polar hyperboloid surface. Thus, it may be possible that, by introducing curvature in the nanorings fabrication, in order to obtain ``hyperboloidal nanorings'', vortices are more stable, and one could fabricate smaller nanomagnets with a vortex as the magnetization groundstate. However, when we think in nanomagnetism applications, this fact does not ensure that nanomagnets with this geometry could support a vortex magnetization configuration with more stability than cylindrical nanorings. To state that, one must consider the volume of a hyperbolic nanoring (nanomagnet limited by two polar hyperboloids with internal and external radii given by $\rho_{1}$ and $\rho_2$, respectively) and take into account the magnetostatic energy to calculate the magnetic energy associated to other magnetization configurations. In this way, if miniaturization of magnetic elements is demanded, the understanding of the curvature influence on magnetic properties of nanostructures is very important, once it may give a way to diminish the size of nanomagnets that present a vortex as the magnetization ground state. In this paper, we have explicitly considered only the vortex state on non-simply connected and negatively curved surfaces by assuming that they satisfy the conditions in their dimensions ensuring this configuration.

Unlike the cylinder surface, the anisotropy parameter presented by the catenoid and hyperboloid, which ensure the vortex as the lower energy configuration, varies with the height and radius of the surface. The results show that the increasing in the catenoid radius produces a decreasing in the modulus of the anisotropy parameter, which goes to 1 at the limit $\rho\rightarrow\infty$, reflecting the fact that, in this limit, the geometry of the catenoid is similar to the cylinder one. The modulus of the anisotropy parameter on the hyperboloid presents a different behaviour, increasing with the surface's radius and height.

Despite we have given attention to magnetic systems, the adopted model may be relevant to study nonmagnetic systems like a superfluid helium film, where $\Phi$ would be the phase of the superfluid order parameter \cite{Vitelli-PRL93}, a superconducting film, where the field is identified with the phase of the collective wave function \cite{Vitelli-PRL93} or nematic liquid crystals, where $\Phi$ would be the nematic director \cite{Napoli-PRL-108}. 

The results obtained here also may be relevant in the magnetoelastic membrane manipulation subject, once vortex configuration could be used as a way for deforming the geometry of magnetoelastic surfaces to obtain a particular shape. In this context, it would be an interesting task to study the vortex energy behaviour on surfaces with positive Gaussian curvature and include the surface tension term in the analysis of this problem. In addition, it may be important to study the influence of an external magnetic field on the energy calculations of magnetoelastic materials presenting a vortex configuration.

\begin{center}
\textbf{Acknowledgements}
\end{center}

We thank the Brazilian agencies CNPq, FAPEMIG and PROPES of the IF Baiano, for financial support. We also thank E.S. Palitot, J.S. Santos, J.M. Fonseca, W.A. Moura-Melo and A.R. Pereira for fruitful discussions. Carvalho-Santos thanks G.H. Lima-Santos and P.G. Lima-Santos for their patience and understanding.\\

\end{document}